\documentstyle[12pt]{article}

\newcommand{\sect}[1]{\setcounter{equation}{0}\section{#1}}
\newcommand{\subsect}[1]{\subsection{#1}}

\def\bq{\begin{equation}}
\def\eq{\end{equation}}
\def\bea{\begin{eqnarray}}
\def\eea{\end{eqnarray}}

\def\1{\'{\i}}

\def\om{J}
\def\th{\theta}
\def\vth{\vartheta}
\def\ss{S}

\def\m{\mu}
\def\l{\lambda}
\def\P{P}

\def\zz{{\bf Z}}

\def\ee{\varepsilon}
\def\calS{{\cal S}}
\def\calI{{\cal I}}
\def\R{\rm I\kern-.2em R}

\def\aa#1#2{\alpha^{#1}_{\,\,#2}}
\def\bb#1#2{\beta_{#1}^{\,\,#2}\,}
\def\cc#1#2{\gamma_{#1}^{\,\,\,\,\,#2}}

\def\k{\kappa}

\begin{document}

\thispagestyle{empty}

\ \vspace{2cm}

\begin{center}
{\LARGE{\bf{The general solution of the real }}

{\LARGE{\bf{$\zz_2^{\otimes N}$} graded contractions of $so(N+1)$}}}
\end{center}

\bigskip\bigskip\bigskip

\begin{center}
Francisco J. Herranz${\dagger}$  and Mariano Santander${\ddagger}$
\end{center}

\begin{center}
\em 
{ { {}${\dagger}$ Departamento de F\1sica, Universidad de Burgos}
\\  E-09006, Burgos, Spain}

{ { {}${\ddagger}$ Departamento de F\1sica Te\'orica, Universidad de
Valladolid } \\   E-47011, Valladolid, Spain }
\end{center}
\rm

\bigskip\bigskip\bigskip\bigskip

\begin{abstract} 
The general solution of the graded contraction equations for a 
$\zz_2^{\otimes N}$ grading of the real compact simple Lie algebra
$so(N+1)$  is presented in an explicit way. It turns out to depend on
$2^N-1$ independent real parameters. The structure of the general graded
contractions is displayed  for the low dimensional cases, and kinematical
algebras are shown to appear straightforwardly. The geometrical (or
physical) meaning of the contraction parameters as curvatures is also
analysed; in particular, for kinematical algebras these curvatures are
directly linked to geometrical properties of possible homogeneous
space-times.   \end{abstract}

\newpage 


\sect{Introduction}

Graded contractions of (complex or real) Lie algebras have been introduced 
by de Montigny, Patera and Moody \cite{MonPat,MooPat} as a new approach
encompassing the study of ordinary contractions of Lie algebras and
allowing the contraction of representations to be simultaneously studied.
The approach is based on the so-called {\em contraction equations}, which
determine all possible {\em contracted} Lie algebras compatible with a
given grading of some initial Lie algebra \cite{PatZas}. Ordinary (simple)
In\"on\"u-Wigner (IW) contractions \cite{IW} appear as related to a $\zz_2$
grading; the solution of the contraction equations is straightforward in
this case. For more  complicated grading groups, and in the complex case,
these equations have been solved for several comparatively small grading
groups (e.g.\ the complete list for $\zz_2,\,  \zz_2 \otimes \zz_2, \,
\zz_3$ is given in \cite{MonPat}), and a computer programme has been
devised for handling more complicated cases \cite{BerMon}.

In a previous paper the graded contractions of the {\em real} orthogonal 
algebra $so(N+1)$ associated to a fine grading group $\zz_2^{\otimes N}$
were studied without relying on a computer programme, and a {\em
particular} set of solutions which depends on $N$ real parameters was given
\cite{HMOS}. To know the general solution of the contraction equations for
this grading would be interesting as a first step to study similar graded
contractions  for algebras in the unitary $su(N+1)$ and symplectic series
$sp(N+1)$, for which a natural $\zz_2^{\otimes N}$ grading can be
``derived" from the orthogonal one \cite{Tesis}; the general solution of
the orthogonal contraction equations could go a long way to provide a
general solution for these other cases. 

In this paper we advance the {\em general} solution for a fine
$\zz_2^{\otimes N}$ grading of the real Lie algebra $so(N+1)$. It should be
recalled that the general solution for a given graded algebra with grading
group $\Gamma$ is some subset of the corresponding list of solutions of the
contraction equations for $\Gamma$, which generically depend only on the
group $\Gamma$. Usually this is a {\em proper} subset; this is due to the
possible presence of {\em irrelevant} contraction parameters which actually
do not appear in the contractions equations for the given algebra (for
instance, a given grading group element may have not a proper associated
subspace, or two graded subspaces may commute in the initial Lie algebra).
As far as we know, the list of generic solutions to the contraction
equations for a $\zz_2^{\otimes N}$ grading group is not known. However,
the strategy of solving generically the contraction equations can be
succesfully bypassed in specific cases, as the example we are about to
discuss clearly shows. 

The paper is organized as follows. The structure of the $\zz_2^{\otimes
N}$ grading of $so(N+1)$  and the corresponding contraction equations are
presented  in the next section.  We solve the contraction
equations in section 3  showing how the $3 {{N+1}\choose{3}}$ initial 
relevant  contraction parameters turn out to  depend on $2^N-1$ {\em
independent} real parameters. All graded contractions are continuous in
this case, and the number of possible contractions equals the expected
number for a general composition of simple IW contractions; the result is
in accordance with the not entirely obvious result in \cite{Weimar2}: any
continuous graded contraction is equivalent to some generalized IW
contraction.  The solution given in \cite{HMOS} appears as a rather
particular case, as it corresponds to having all but $N$ parameters fixed
to 1. Explicit results for the simplest cases with $N=2,3$ are  given in
section 4 in order to clearly describe the structure of the general
solution; moreover for $N=3$ we introduce all the (2+1) kinematical
algebras \cite{BLL} within the graded contracted algebras of $so(4)$ in a
straightforward way, thus giving derivation of Lie algebra kinematics from
the graded contraction perspective alternative to the one discussed in
\cite{Tolar}. An interesting byproduct in this approach is the
interpretation of the contraction coefficients as related to curvatures of
homogeneous spaces.


\sect{The contraction equations}

Recall that a {\em grading} of a real Lie algebra $L$ by an Abelian
finite group $ \Gamma$ \cite{PatZas} is a decomposition of the vector space
structure of $L$:
\bq
L=\bigoplus_{\mu\in  \Gamma} L_\mu ,  
\label{bba}
\eq
such that if $x \in L_\mu$ and $y \in L_\nu$ then $[x,y]$ belongs to
$L_{\mu+\nu}$:
\bq
[L_\mu,L_\nu] \subseteq L_{\mu+\nu}, \qquad
\mu, \, \nu, \, \mu+\nu\in  \Gamma. 
\label{bbb} 
\eq

A (real) {\em graded contraction} of the real Lie algebra
$L$ \cite{MonPat,MooPat} is a real Lie algebra   $L_\ee$  with the same
vector space structure  as $L$, but with Lie brackets for $x\in L_\mu$ and
$y\in L_\nu$ modified as follows:   
\bq 
[x,y]_\ee := \ee_{\mu,\nu} \,[x,y], \quad {\mbox{in short hand form}}
\quad [ L_\mu, L_\nu ]_\ee := \ee_{\mu,\nu} \, [ L_\mu, L_\nu],
\label{bbc} 
\eq
where the {\em contraction parameters} $\ee_{\mu,\nu}$ are real numbers
such that $L_\ee$ is a Lie algebra; they must satisfy the {\em contraction
equations}:  
\bq
\ee_{\mu,\nu} = \ee_{\nu,\mu}  \qquad \qquad 
\ee_{\mu,\nu} \, \ee_{\mu+\nu,\sigma}=
\ee_{\mu,\nu+\sigma} \, \ee_{\nu,\sigma} 
\label{bbe}  
\eq 

The Lie algebra $so(N+1)$ has $N(N+1)/2$ generators $\om_{ab}$
$(a,b=0,1,\dots, N, \  a<b)$ with non-zero Lie brackets: 
\bq
[\om_{ab}, \om_{ac}] =  \om_{bc}, \qquad
[\om_{ab}, \om_{bc}] = -\om_{ac}, \qquad
[\om_{ac}, \om_{bc}] =  \om_{ab}, \qquad  a<b<c .
\label{bbz}  
\eq
 
The fine grading group $\Gamma$ of $so(N+1)$ we are going to deal with is
isomorphic to  $\zz_2^{\otimes N}$ and  is generated by a set of $2^N$ 
commuting involutive automorphisms $S_\calS: so(N+1)\to so(N+1)$ where
$\calS$ is any subset of the set of indices $\calI=\{0,1,\dots,N\}$ (see
\cite{HMOS} for more details). The automorphism $S_\calS$ is defined as
\bq
\ss_{\calS} \om_{ab} = (-1)^{\chi_{\calS}(a)+\chi_{\calS}(b)}\om_{ab},
\label{bbf}  
\eq
where $\chi_{\calS}(i)$ is the characteristic function over $\calS$, which
equals either $1$ or $0$ according as $i \in \calS$ or $i \notin \calS$.  

Each involutive automorphism  $S_\calS$ provides a $\zz_2$ grading of
$so(N+1)=L_0\oplus L_1$ where $L_0$ is the $S_\calS$-invariant subspace
(spanned by the generators $\om_{ab}$ with either both indices or none in 
$S_\calS$), while $L_1$ is the $S_\calS$-anti-invariant subspace (spanned
by  the  $\om_{ab}$ with a single index  in  $S_\calS$). Note also that
$\ss_{\calS} \equiv \ss_{\calI\backslash\calS}$, i.e.\ the automorphism
associated to a subset $\calS$ is the same as the one associated to the
complement  ${\calI\backslash\calS}$ in the whole set of indices $\calI$.
For instance, if $\calI=\{0,1,2,3\}$, then $S_{012}\equiv S_3$, 
$S_{13}\equiv S_{02}$, etc.

We choose the $N$ automorphisms $ \ss_{{0}}, \  \ss_{{01}},\  \ss_{{012}},  
\dots ,\  \ss_{{01\dots N-1}}$ as a basis for the Abelian grading
group $\Gamma$ of $so(N+1)$. Thus, a generic element $\mu\in \Gamma$ can be
written as 
\bq
\mu=\prod_{k=0}^{N-1} (\ss_{{01\dots k}})^{\mu_k},\qquad 
\mu_k \in \{0,1\}.
\label{aa}
\eq
 A generator $J_{ab}$ of $so(N+1)$ belongs to the (one-dimensional) 
grading subspace $L_\mu$ where the sequence $\mu_k$ of ``coordinates" of
$\mu$ is characterized by a  contiguous string of 1's  starting at the
$a$th position and ending at the $(b-1)$th position with 0's at the
remaining places:   \bq
\langle\om_{ab} \rangle = L_\mu \quad \equiv \quad 
\mu = \{ {0\dots 0 1_a\dots 1 0_b\dots 0} \}. 
\label{bbi}  
\eq 
Therefore we can denote each particular $\mu$ actually appearing in
the decomposition (\ref{bba}) for the specific grading we are dealing with  
by the pair of indices
$\mu
\equiv ab,
\  (a<b)$ instead of using its complete string. The contraction parameters
$\ee_{\mu,\nu}$ where at least one of $\mu$, $\nu$ and $\mu+\nu$ is not of
the form (\ref{bbi}) are therefore irrelevant, as the corresponding grading
subspace is the trivial null subspace. On the contrary, we call {\em
relevant}  contraction parameters those appearing in the contractions
equations; they must have $\mu$, $\nu$ and $\mu+\nu$ of the form
(\ref{bbi}). In \cite{HMOS} it was shown that in this case there are ${3}
{N+1\choose 3}$  {\em relevant} contraction parameters. They can be
classified in three disjoint sets:    \bq 
\aa{a}{bc}  \equiv \ee_{ab,ac}, \qquad 
\bb{ac}{b}  \equiv \ee_{ab,bc}, \qquad 
\cc{ab}{c}  \equiv \ee_{ac,bc}, \qquad a<b<c .
\label{bbj}  
\eq 
All contraction equations coming from (\ref{bbe}) are naturally classed
into groups of 12 equations, one group for each ordered set of {\em four}
indices $a<b<c<d$:
\bea
\bb{ac}{b} \bb{ad}{c}  = \bb{ad}{b} \bb{bd}{c} \ \ & &      \cr
\aa{a}{bc} \bb{bd}{c}  = \aa{a}{bd} \bb{ad}{c} \ \  && 
\aa{a}{bd} \aa{b}{cd}  = \aa{a}{cd} \bb{ac}{b} \qquad  
\aa{a}{bc} \aa{b}{cd}  = \aa{a}{cd} \bb{ad}{b}              \cr
\aa{a}{cd} \cc{bc}{d}  = \aa{a}{bc} \cc{ab}{d}  \ \  &&
\aa{a}{bd} \cc{bc}{d}  = \aa{a}{bc} \cc{ac}{d}              \label{bbk}\\
\bb{ad}{b} \cc{ac}{d}  = \bb{ac}{b} \cc{bc}{d}  \ \  &&
\bb{bd}{c} \cc{ab}{d}  = \cc{ab}{c} \cc{ac}{d} \qquad \ 
\bb{ad}{c} \cc{ab}{d}  = \cc{ab}{c} \cc{bc}{d}               \cr
\aa{a}{cd} \bb{bd}{c}  = \aa{a}{bd} \cc{ab}{c}  \ \  &&
\aa{b}{cd} \bb{ad}{c}  = \bb{ad}{b} \cc{ab}{c} \qquad \ \ 
\aa{b}{cd} \cc{ac}{d}  = \bb{ac}{b} \cc{ab}{d} 
\nonumber
\eea
In \cite{HMOS} the solution of these equations under the condition
 $\bb{ac}{b}\ne 0$ was derived; each such solution turns out to be
equivalent to a solution with all $\bb{ac}{b}=1$ and then the equations
(\ref{bbk}) dramatically simplify, so that all contraction parameters can
be expressed in terms of $N$ real independent parameters.  However, when
some $\bb{ac}{b}$ are allowed to be equal to zero, the equations are rather
complicated, and the naive case-by-case analysis succesfully done for
smaller grading groups is quickly realised an unfeasible.


\sect{The general solution}

Consider real functions $\th$ defined on the collection of all subsets
of $\calI=\{0,1,\dots,N\}$, $\th : {\cal P}(\calI) \longrightarrow \R$, 
satisfying the additional condition
$\th({\calS}) \equiv \th({\calI\backslash \calS})$. We denote the common
value $\th({\calS}) \equiv \th({\calI\backslash \calS})$ as 
$\th_{\calS}^{\calI\backslash \calS} \equiv \th_{\calI\backslash
\calS}^{\calS}$; there are $2^N$ such values, one of which is 
$\th_{\cal I}^{\emptyset}$. The general solution of the system (\ref{bbk}) 
can be expressed in terms of these values, taken as independent parameters,
according to following statement:

\noindent {\bf Theorem} 
\begin{em}
The general solution of the $\zz_2^{\otimes N}$ graded contractions of the
Lie algebra ${so}(N+1)$ depends on $2^{N}-1$ real independent parameters  
$\th_{\calS}^{\calI\backslash  \calS}$ where $\calI=\{0,1,\dots,N\}$
and $\cal S$ is a proper subset of $\cal I$: $\calS\subset \calI$. The
relevant contraction parameters are given by:
\bea
&& \aa{a}{bc} = \prod_{\calS}\th_{\calS}^{\calI\backslash  \calS } 
              = \prod \th_{bc\dots}^{a\dots},                   \quad 
{ \mbox{with}}\ \{b,c\}\subseteq \calS\ { \mbox{and}}\  
\{a\}\notin\calS; \cr
&& \bb{ac}{b} = \prod_{\calS}\th_{\calS}^{\calI\backslash  \calS }
              = \prod \th_{ac\dots}^{b\dots},\quad\ \ \!
{\mbox{with}}\  \{a,c\}\subseteq \calS\ { \mbox{and}} \ \{b\}\notin \calS;
\label{aad}    \\
&& \cc{ab}{c} = \prod_{\calS}\th_{\calS}^{\calI\backslash \calS } 
              = \prod \th_{ab\dots}^{c\dots},       \quad 
{ \mbox{with}}\ \{a,b\}\subseteq\calS\  { \mbox{and}}\  \{c\}\notin \calS,
\nonumber
\eea
where the products with index $\calS$ run over all possible (proper)
subsets of $\calI$ that satisfy the conditions imposed in each case. The
non-identically zero Lie brackets of the contracted Lie algebra obtained
from ${so}(N+1)$ are
\bq
[\om_{ab}, \om_{ac}] =  \aa{a}{bc} \om_{bc}, \quad
[\om_{ab}, \om_{bc}] = -\bb{ac}{b} \om_{ac}, \quad
[\om_{ac}, \om_{bc}] =  \cc{ab}{c} \om_{ab}, \quad  a<b<c,
\label{aae}
\eq
without sum over repeated indices and with $\aa{a}{bc}$, $\bb{ac}{b}$, 
$\cc{ab}{c}$ given by (\ref{aad}).
\end{em}
 
\noindent
{\em Proof}. The proof is rather direct but somewhat tedious; we restrict
here to comment the main lines.  Each of the contraction equations given in
(\ref{bbk}) is like $MN=PQ$, where each term carries three indices (two
subindices and a third single superindex), taken out of four. The general
solution of each such equation can be given in terms of eight parameters, 
$m^1, m_1, n^1, n_1, p^1, p_1, q^1, q_1$ as
\bq
M=m^1 m_1,\qquad N=n^1 n_1,\qquad P=p^1 p_1,\qquad Q=q^1 q_1,
\label{ma}
\eq
which are however not independent, but must be subjected to four auxiliary
relations
\bq
m^1 = p^1,\qquad  m_1 = q_1,\qquad n^1 = q^1,\qquad n_1 = p_1.
\label{mb}
\eq
 Now repeat this
decomposition for each equation (\ref{bbk}), writing each parameter $m$,
(resp.\ $n, p, q$) as a symbol $\vth$ with two groups of indices, the
first one with the same index structure as $M$, (resp.\ $N, P, Q$) and
taking for the second group the fourth index already present in the
equation but not in $M$ (resp.\ $N, P, Q$), placed either as a superindex
or as a subindex, instead of the index $1$ above. For instance,  the first
equation of (\ref{bbk}) would lead to
\bq
\bb{ac}{b} = \vth_{ac,}^{\ b\,,d} \, \vth_{ac,d}^{\ b\,,}\qquad
\bb{ad}{c} = \vth_{ad,}^{\ c\,,b} \, \vth_{ad,b}^{\ c\,,}\qquad
\bb{ad}{b} = \vth_{ad,}^{\ b\,,c} \, \vth_{ad,c}^{\ b\,,}\qquad
\bb{bd}{c} = \vth_{bd,}^{\ c\,,a} \, \vth_{bd,a}^{\ c\,,}
\label{mc}
\eq
with the $\vth$ symbols satisfying 
\bq
\vth_{ac,}^{\ b\,,d}= \vth_{ad,}^{\ b\,,c}\qquad
\vth_{ac,d}^{\ b\,,}=\vth_{bd,a}^{\ c\,,}\qquad
\vth_{ad,}^{\ c\,,b}=\vth_{bd,}^{\ c\,,a}\qquad
\vth_{ad,b}^{\ c\,,}=\vth_{ad,c}^{\ b\,,}  
\label{md}
\eq

As long as all auxiliary relations are satisfied, this transforms all
contraction equations into identities, at the expense of introducing a
rather large number of parameters, which are however subjected to a number
of auxiliary relations (similar to (\ref{md})), which can be then
eliminated in some adequate way. The result of the elimination boils down
to two simple rules. First, each symbol 
$\vth_{ac,}^{\ b\,,d}$, $\vth_{ac,d}^{\ b\,,}$, \dots actually depends on
the two subsets of $\cal I$ made up with the union of all subindices and
the union  of all superindices, so that e.g.\ $\vth_{ab,c}^{\ d,} =
\vth_{ac,b}^{\ d\,,} = \vth_{bc,a}^{\ d,}$ which will be simply denoted
$\th_{abc}^{d}$, likewise,    \ $\vth_{ac,}^{\ d,b} = \vth_{ac,}^{\
b\,,d}$  will be denoted $\th_{ac}^{bd}$. Second, each symbol
$\th_{ac}^{bd}$ depends only on the two subsets of indices, but not on
their position as subindices or superindices, so that e.g.\
$\th_{ac}^{bd}=\th_{bd}^{ac}$.   

However, these four index $\th$ symbols are not independent. To see this,
recall that each contraction equation involves four contraction
coefficients. By using the previous device, each of these coefficients can
be written as the product of two {\em four} index $\th$ symbols. But each
contraction coefficient appears several times in the whole set of
contraction equations, and to each appearance a decomposition as a product
of two $\th$ symbols with four indices has been allocated. For instance,
the coefficient $\bb{36}{5}$ will appear in the first equation (\ref{bbk})
for $abcd=1356$ and also for $abcd=3456$. For each of these appearances we
have 
\bq
\bb{36}{5}=\vth_{36,}^{\ 5\,, 1} \vth_{36,1}^{\ 5\,,} 
=\th_{36}^{15}\th_{136}^{5}
\qquad
\bb{36}{5}=\vth_{36,}^{\ 5\,, 4} \vth_{36,4}^{\ 5\,,} 
=\th_{36}^{45}\th_{346}^{5}
\label{mmd}
\eq
so the  coefficients $\th$ must satisfy
\bq
\th_{36}^{15}\th_{136}^{5}=\th_{36}^{45}\th_{346}^{5}
\label{nnd}
\eq
In the same way we will have  to
enforce the equality of many such products. This turns out in a number of
quadratic equations, whose structure is again similar to the initial
equations, but each involving {\em five} indices taken out of the set
$\calI$, so the same decomposition procedure can be applied again. For
instance, each coefficient $\th$ in the equation (\ref{nnd}) would be
decomposed as a product of two five index $\vth$ symbols, e.g.,  
$\th_{36}^{15}=\vth_{36,}^{15\,, 4} \vth_{36,4}^{1\,,5}$, etc, where the
set of five index coefficients $\vth$ must satisfy auxiliary
equations, derived again from (\ref{mb}) and  similar to (\ref{md}). The
elimination of these auxiliary equations boils down to the same simple
rules stated before  (e.g., $\vth_{36,}^{15\,, 4}=\vth_{36,}^{14\,,
5}=\dots $, which will be denoted $\th_{36}^{145}$, etc. and 
$\th_{36}^{145}=\th_{145}^{36}$, etc.). Now all equations like 
(\ref{nnd}) are turned into identities, and only the auxiliar equations for
the five index $\th$ symbols remain.

The process is iterated until no more indices are left, at which point all
equations are transformed into identities; this explains the structure of
the solution. It is easy to check that after using (\ref{aad}) all the
contraction equations (\ref{bbk}) are turned into identities. 
 
\medskip

It is worth remarking that there exists a close relationship between
the parameters $\th_{\calS}^{\calI\backslash\calS}$ and the involutive
automorphisms  $\ss_\calS$. Each non trivial involution $\ss_\calS$ gives
rise to a simple IW contraction whose effect consists on  a graded scale
change   with scaling factor  $\lambda$ on the anti-invariant generators
under $\ss_\calS$ (those  $\om_{ab}$ where either $a$ or $b$ belongs to 
$\calS$) followed by the limit $\lambda\to 0$. This scale change only
modifies the parameter $\th_{\calS}^{\calI\backslash\calS}
\to \lambda^2\th_{\calS}^{\calI\backslash\calS}$, the remaining ones being 
invariant,  and in the limit $\lambda\to 0$ this parameter vanishes. Thus,
there are $2^N-1$  simple IW contractions associated to the same number of
non trivial involutions or of $\zz_2$ subgradings; the identity
involution $\ss_\calI$ would be associated with $\th_{01\dots
N}^{\emptyset}$, which is the only value of 
$\th_{\calS}^{\calI\backslash\calS}$ not appearing explicitly in
(\ref{aad}). The composition of two or more of such contractions is a
generalized IW contraction \cite{Weimar2} where more than one parameter
$\th_{\calS}^{\calI\backslash\calS}$ go to zero at the same time (with
possibly different powers of $\lambda$). 

It is also clear that all graded contractions are continuous for the
grading we are dealing with, as the identity element in the grading group
has no an associated  proper subspace. 

The graded contractions of ${so}(N+1)$ with all
$\th_{\calS}^{\calI\backslash  \calS} \neq 0$  give  rise  to the different
non-compact real forms   ${so}(p,q)$ with $p+q=N+1$  (besides the original
${so}(N+1)$). In this case  the graded contraction is not a contraction of
the initial algebra in its original meaning of limiting process. When one
or more $\th_{\calS}^{\calI\backslash \calS} $ are zero,  a non simple Lie
algebra is obtained.  It is interesting to note that the whole family of
graded contractions of $so(N+1)$ affords a sort of ordered ``lattice" of
algebras, starting at the simple real algebras $so(p,q)$ and ending at the
extreme case, when all $\th_{\calS}^{\calI\backslash  \calS}=0 $, into the
Abelian algebra with all commutators zero.

If we only  allow the $N$ parameters $\th_{0}^{12\dots N}$,  
$\th_{01}^{23\dots N}$, $\th_{012}^{34\dots N},\dots, 
\th_{012\dots N-1}^{N}$ 
to take over arbitrary values enforcing the  value 1 for all the remaining
ones, the contracted algebra commutation relations are:
\bq
[\om_{ab}, \om_{ac}] = \k_{ab}\, \om_{bc}, \quad
[\om_{ab}, \om_{bc}] = -  \om_{ac}, \quad
[\om_{ac}, \om_{bc}] = \k_{bc}\, \om_{ab}, \quad  a<b<c ;
\eq
where the $\k$ coefficients are   defined as:
\bea
&&\k_{ab} := \k_{a+1} \k_{a+2} \dots \k_b, \qquad a,b=0,\dots,N; 
\quad a<b,\\
&&\k_{a}:= \th_{0\dots a-1}^{a\dots N},\qquad a=1,\dots,N.
\eea

In this way the so-called {\em Cayley--Klein algebras}, which were the
particular case studied in \cite{HMOS}, are recovered. As a collective,
this subfamily of graded contractions inherits (in a more complicated form)
most properties coming from the simple nature of the algebras $so(p,q)$,
and is termed {\em quasisimple} in the literature \cite{Rosenfeld}.


\sect{Examples}

Let us illustrate the results of the theorem for $so(3)$ and $so(4)$,
where  from the two ways of writing each coefficient $\th$, we  have chosen
the one  where $0$ appears as a subindex, even if this sometimes aparently
spoils the simplicity of the rule (\ref{aad}).

\subsect{$so(3)$}
 
The grading group is $\zz_2\otimes \zz_2$ and is generated by the
automorphisms $\ss_{{0}}$ and  $\ss_{{01}}$ acting on the generators
$\{\om_{01}, \om_{02}, \om_{12}\}$ as:
\bea
&& \ss_{{0}} : (\om_{01},\om_{02},\om_{12}) \longrightarrow
               (-\om_{01},-\om_{02},\om_{12}),                \cr
&& \ss_{{01}}: (\om_{01},\om_{02},\om_{12}) \longrightarrow
               (\om_{01},-\om_{02},-\om_{12}).
\label{aaf}
\eea
These involutions endow the basis of   ${so}(3)$ with the following grading:
\bq
L_{\{01\}} \equiv L_{12} = \langle \om_{12}\rangle , \quad
L_{\{10\}} \equiv L_{01} = \langle \om_{01}\rangle , \quad
L_{\{11\}} \equiv L_{02} = \langle \om_{02}\rangle ,  
\label{aag}
\eq
where the indices between brackets in $L_\mu$ denote the whole sequence
$\{\mu_k\}$. There are ${3} {2+1\choose3}=3$ relevant contraction
coefficients (one $\alpha$, one $\beta$ and one  $\gamma$) which
depend on $2^2-1=3$ parameters
$\th_{0}^{12}\equiv\th_{12}^{0}, \ \th_{02}^{1}\equiv\th_{1}^{02}$ and
$\th_{01}^{2}\equiv\th_{2}^{01}$: 
\bea
&& \ee_{\{10\},\{11\}} = \ee_{01,02} \equiv
                          \aa{0}{12}=\th_{0}^{12},   \cr
&& \ee_{\{10\},\{01\}} = \ee_{01,12} \equiv
                          \bb{02}{1}=\th_{02}^{1},    \label{aah} \\
&& \ee_{\{01\},\{11\}} = \ee_{02,12} \equiv 
                          \cc{01}{2}=\th_{01}^{2}. 
\nonumber
\eea

The commutation relations for the contracted algebra of ${so}(3)$ are:
\bq
[\om_{01},\om_{02}] =  \th_{0}^{12} \om_{12}, \quad
[\om_{01},\om_{12}] = -\th_{02}^{1} \om_{02}, \quad 
[\om_{02},\om_{12}] =  \th_{01}^{2} \om_{01} .   
\label{aai}
\eq
This family of algebras includes ${so}(3)$, ${so}(2,1)$,
the Euclidean ${e}(2)$, Poincar\'e ${p}(1+1)$, Galilean ${g}(1+1)$
and the Abelian algebras. Upon graded contraction equivalence, the value
$\th_{02}^{1}$ can be reduced either to $0$ or $1$, and then each of the
two remaining  contraction parameters $\th_{0}^{12}$ and $\th_{01}^{2}$ can
be reduced to either $1$, $0$ or $-1$. 

This example allows to see clearly the point commented upon in the
introduction: acording to the results in \cite{MonPat}, for a  generic
$\zz_2 \otimes \zz_2$ graded structure, there exists $40$ non-equivalent
solutions of the complex graded equations. Even if we are dealing with
real graded contractions, the number of inequivalent solutions here is
much lesser. The reason is easy to see: for the algebra $so(3)$ and the
grading (\ref{aag}), the subspace $L_{\{00\}}$ is the trivial null
subspace, so many contraction parameters which are relevant in the generic
case, turn out to be irrelevant here. From another point of view, this case
has been also discussed in \cite{Weimar1}. 

It is worthy to analyze the meaning of the three contractions constants in
this example; in fact the most interesting traits of the $N$d case are
already present in this simplest case. First, a notation
change helps in the interpretation: we shall rewrite (\ref{aai}) as
\bq
[\P_1,\P_2] =  \m_1 J, \quad
[\P_1,J] = -\l \P_2, \quad 
[\P_2,J] =  \m_2 \P_1.   
\label{aaii}
\eq
Now for each algebra (\ref{aaii}) we can build three two-dimensional
symmetrical homogeneous spaces, each associated to the involutions $
\ss_{{0}}, \  \ss_{{01}},  \  \ss_{{02}}$, taking the coset space of the
graded contrated group by the subgroup generated by the elements invariant
under the involution. The first two spaces are called the space of points,
and the space of lines. Each of these spaces has a canonical connection, as
well as a compatible canonical (hierarchy of) metrics, coming from a
suitably modified ``Cartan-Killing" form, which is defined even for
non-semisimple cases and reduces to the standard one for $so(3)$ and
$so(2,1)$ \cite{Tesis}. Then the constants $\m_1$ and $\m_2$ turn out to be
equal to the canonical curvature of the spaces of points and lines. The
constant $\l$ plays a similar role for the third homogeneous space
corresponding to the involution $\ss_{{02}}$ (the space of second kind
lines).  

\subsect{$so(4)$}

Let us consider now the $N=3$ case, with basis
$\{\om_{01}, \om_{02}, \om_{03}, \om_{12}, \om_{13}, \om_{23}\}$.
The group $\zz_2 ^{\otimes 3}$ determines the graded subspaces:
\bea
&& L_{\{100\}} \equiv L_{01} = \langle \om_{01} \rangle, \quad
   L_{\{110\}} \equiv L_{02} = \langle \om_{02} \rangle, \quad
   L_{\{111\}} \equiv L_{03} = \langle \om_{03} \rangle,         \cr
&& L_{\{010\}} \equiv L_{12} = \langle \om_{12} \rangle, \quad
   L_{\{011\}} \equiv L_{13} = \langle \om_{13} \rangle, \quad
   L_{\{001\}} \equiv L_{23} = \langle \om_{23} \rangle.  
\label{aaj}
\eea
There are $3 {{3+1}\choose 3} = 12$ relevant contraction parameters which 
can be written in terms of $2^3-1=7$ coefficients 
$\th_{0}^{123}$, $\th_{01}^{23}$,  $\th_{02}^{13}$, $\th_{03}^{12}$,
$\th_{012}^{3}$,  $\th_{013}^{2}$, $\th_{023}^{1}$ as follows:  
\bea
&& \ee_{\{100\},\{110\}} = \ee_{01,02} \equiv \aa{0}{12}
                         = \th_{0}^{123}\th_{03}^{12}     \qquad
   \ee_{\{100\},\{111\}} = \ee_{01,03} \equiv \aa{0}{13}
                         = \th_{0}^{123}\th_{02}^{13}              \cr
&& \ee_{\{110\},\{111\}} = \ee_{02,03} \equiv \aa{0}{23}
                         = \th_{0}^{123}\th_{01}^{23}     \qquad
   \ee_{\{010\},\{011\}} = \ee_{12,13} \equiv \aa{1}{23}
                         = \th_{01}^{23}\th_{023}^{1}              \cr
&& \ee_{\{010\},\{100\}} = \ee_{01,12} \equiv \bb{02}{1}
                         = \th_{02}^{13}\th_{023}^{1}     \qquad\ \ 
   \ee_{\{011\},\{100\}} = \ee_{01,13} \equiv \bb{03}{1}
                         = \th_{03}^{12}\th_{023}^{1}              \cr
&& \ee_{\{001\},\{110\}} = \ee_{02,23} \equiv \bb{03}{2}
                         = \th_{03}^{12}\th_{013}^{2}     \qquad\ \ 
   \ee_{\{001\},\{010\}} = \ee_{12,23} \equiv \bb{13}{2}
                         = \th_{02}^{13}\th_{013}^{2}              \cr
&& \ee_{\{110\},\{010\}} = \ee_{02,12} \equiv \cc{01}{2}
                         = \th_{01}^{23}\th_{013}^{2}      \qquad
   \ee_{\{111\},\{011\}} = \ee_{03,13} \equiv \cc{01}{3}
                         = \th_{01}^{23}\th_{012}^{3}             \cr
&& \ee_{\{111\},\{001\}} = \ee_{03,23} \equiv \cc{02}{3}
                         = \th_{02}^{13}\th_{012}^{3}     \qquad
   \ee_{\{011\},\{001\}} = \ee_{13,23} \equiv \cc{12}{3}
                         = \th_{03}^{12}\th_{012}^{3}              \cr
&& 
\eea
and for the contracted algebra we have the following non-identically
vanishing Lie brackets:
\bea
&&[\om_{01}, \om_{02}]  = \th_{0}^{123}\th_{03}^{12} \om_{12}\qquad
[\om_{01}, \om_{12}]   = -\th_{02}^{13}\th_{023}^{1}\om_{02} \qquad
 [\om_{02}, \om_{12}]   =\th_{01}^{23}\th_{013}^{2} \om_{01} \cr
&&[\om_{01}, \om_{03}]  = \th_{0}^{123}\th_{02}^{13} \om_{13}  \qquad
 [\om_{01}, \om_{13}]  = -\th_{03}^{12}\th_{023}^{1}\om_{03} \qquad
[\om_{03}, \om_{13}]  = \th_{01}^{23}\th_{012}^{3} \om_{01}\cr
&&[\om_{02}, \om_{03}]  = \th_{0}^{123}\th_{01}^{23} \om_{23}\qquad
 [\om_{02}, \om_{23}]  = -\th_{03}^{12}\th_{013}^{2} \om_{03}\qquad
[\om_{03}, \om_{23}]  = \th_{02}^{13}\th_{012}^{3} \om_{02}\cr
&&[\om_{12}, \om_{13}]  = \th_{01}^{23}\th_{023}^{1} \om_{23} \qquad
[\om_{12}, \om_{23}]  = -\th_{02}^{13}\th_{013}^{2} \om_{13}\qquad
[\om_{13}, \om_{23}]  = \th_{03}^{12}\th_{012}^{3} \om_{12}\cr
&& \label{aak}
\eea
For each three dimensional subalgebra, a structure like (\ref{aai}) is
found, the structure constants being now products of two parameters $\th$.

It is interesting  to find out how the eleven (2+1) kinematical algebras
\cite{BLL} appear as particular graded contracted algebras of $so(4)$.
Let us take a physical basis $\{H,P_1,P_2,K_1,K_2,J_3\}$ whose elements
generate the time translation, two space translations, two boosts and one
space rotation, respectively. We can consider the following identification
with the ``abstract" basis
$\{\om_{ab}\}$:
\bq
H\equiv \om_{01},\quad 
P_1\equiv \om_{02},\quad 
P_2\equiv \om_{03},\quad 
K_1\equiv \om_{12},\quad 
K_2\equiv \om_{13},\quad 
J_3\equiv \om_{23} .
\label{aal}
\eq

The most restrictive requirement to be imposed on these algebras is the
automorphism condition on parity and time reversal; this is automatically
taken into account using graded contractions, and these transformations
correspond to the automorphisms $\ss_{{01}}$ and $\ss_{{023}}$ respectively. 
Ordinary physical space isotropy is translated into the requirement
\bq
[J_3,H]=0,\qquad [J_3,P_i]=\epsilon_{3ij}P_j,\qquad
[J_3,K_i]=\epsilon_{3ij}K_j .
\label{aam}
\eq
This condition implies
\bq
\th_{03}^{12}\th_{013}^{2}=\th_{02}^{13}\th_{012}^{3}
=\th_{02}^{13}\th_{013}^{2}=\th_{03}^{12}\th_{012}^{3}=1.
\eq
Then the four coefficients 
  $\th_{02}^{13}$, $\th_{03}^{12}$,
$\th_{012}^{3}$,  $\th_{013}^{2}$ must be simultaneously different from
zero, and are determined by any of them, which by means of a simple
rescaling can be made equal to $1$; we will asume therefore that all these
contraction coefficients rest equal to $1$ (this way euclidean-like  space
isotropy will be preserved under graded contractions). 

So any possible
kinematical algebra is completely described by the values of three {\it
independent}  contraction constants, $\th_{0}^{123}$,
$\th_{01}^{23}$ and  $\th_{023}^{1}$, a rather expected outcome as
imposing space isotropy we are indeed reducing the problem to a (1+1)
kinematics, where the most general solution (\ref{aai}) depends on three
contraction parameters. The commutators (\ref{aak}) in the basis
(\ref{aal}) read: 
\bea 
&&[H, P_i]  = \th_{0}^{123}\, K_i ,\qquad
[H, K_i]  = -\th_{023}^{1}\, P_i,\qquad
[P_i,K_j]=\delta_{ij}\th_{01}^{23}\, H,\cr
&&[P_1, P_2]  = \th_{0}^{123} \th_{01}^{23}\, J_3,\qquad
[K_1, K_2]  = \th_{01}^{23} \th_{023}^{1}\, J_3 ,\qquad i,j=1,2.
\label{aan}
\eea
Finally the boosts generate a non-compact group
when $\th_{01}^{23} \th_{023}^{1}\le 0$, so this condition should be also
enforced. 

A more clear view is obtained by performing now a notational change: 
$\th_{0}^{123} \equiv \m_1$, $\th_{01}^{23} \equiv \m_2$ and  
$\th_{023}^{1}
\equiv \l$, so that the (1+1) kinematical subalgebra generated by $H, P_1,
K_1$ (and also $H, P_2, K_2$) closes a Lie algebra (\ref{aaii}):
\bea
&&[H, P_i]  = \m_1\, K_i ,\qquad
[H, K_i]  = -\l\, P_i,\qquad
[P_i,K_j]=\delta_{ij}\m_2\, H,\cr
&&[P_1, P_2]  = \m_1\m_2\, J_3,\qquad
[K_1, K_2]  = \m_2 \l\, J_3 ,\qquad i,j=1,2.
\label{aao}
\eea

The constant $\l$ can be reduced either to zero or to 1
by means of an equivalence; we will always asume it takes on these values.
The physical meaning of contraction parameters can be also clearly seen in
the commutation relations (\ref{aao}) as linked to geometrical properties
of the corresponding homogeneous spaces. The most important is space-time
itself: space-time curvature (i.e., along 2-flat directions like ($tx$) and
($ty$)) equals to
$\m_1$, so De Sitter and Newton-Hooke universes have non-zero constant
space-time curvature, while Galilei or Minkowski have zero curvature.
Two-dimensional space curvature (along the 2-flat ($xy$), we could say
space-space curvature) equals the product $\m_1 \m_2$, and can therefore be
zero when $\m_2=0$, even if the space-time curvature is different from
zero; this is the case in the non-relativistic Newton-Hooke algebras, which
has a flat $2$-space. However, when $\m_2 \neq 0$, space-time curvature and
space curvature are linked, and appear simultaneously.

Likewise, the constant $\m_2$ is the curvature of the  space of lines,
which is negative in the ``relativistic" space-times and zero in the
non-relativistic ones; positive values are not allowed as they would lead
to compact inertial transformations. In fact the curvature of the space of
lines is linked to the fundamental constant $c$ of relativistic theories as
$\m_2=\frac{-1}{c^2}$, and special relativity is no more than stating that
the kinematical space of lines has a constant non-zero (negative)
curvature. 

Each of the possible values of the pair $\m_2, \l$ is coupled  with the
three possible values for the ordinary space-time curvature
$\m_1$. This way, all graded contraction constants appear as physical
parameters, whose values are determined by the geometrical properties of
space-time itself. 

We summarize the results in  
table I, where the usual physical name of the
algebra is displayed along with the values of the contraction
coefficients. The first six algebras are ``relative-time" type, while the
six remaining are ``absolute-time" type. The space-time, speed-space and
speed-time contractions correspond, in this order, to the cancelation of
$\m_1$, $\m_2$ and  $\l$, and the algebras are classed naturally in four
groups of three algebras, corresponding to the three essentially different
values of $\m_1$; The unphysical para-Galileicase is somewhat exceptional
as both $\m_1=1$ and $\m_1=-1$ lead to isomorphic algebras. 

\bigskip

\noindent
{\bf {Table I}}.  Kinematical algebras as graded contractions of $so(4)$. 
\smallskip

\noindent
\begin{tabular}{|ll|rrr|}
\hline
 \multicolumn{2}{|c|}{Kinematical
algebra}&$\m_1$&$\m_2$&$\l$\\ 
\hline
De Sitter &$so(3,1)$& -1&-1&$1$\\
Poincar\'e &$iso(2,1)$&0&-1&$1$\\
Anti-De Sitter& $so(2,2)$&$1$&-1&$1$\\
Inhomogeneous $so(3)$  & $iso(3)$&-1&-1&0\\
Carroll &$ii'so(2)$&0&-1&0\\
Para-Poincar\'e &$iso(2,1)$&$1$&-1&0\\
\hline
Expanding Newton--Hooke&$t_4(so(2)\oplus so(1,1))$&-1&0&$1$\\
Galilei &$iiso(2)$&0&0&$1$\\
Oscillating Newton--Hooke&$t_4(so(2)\oplus so(2))$&$1$&0&$1$\\
Para-Galilei &$iiso(2)$&-1&0&0\\
Static& &0&0&0\\ 
Para-Galilei &$iiso(2)$&1&0&0\\
\hline
\end{tabular}

\bigskip

The kinematical algebras in 
 higher dimensions can be obtained in a similar way.
 We would  also like to recall that a family of
$\zz_2 \otimes \zz_2$ graded contractions of any real form of the complex
Lie algebra $B_2$ contains the (3+1) kinematical algebras; this procedure
was used in \cite{Tolar}.



\bigskip
\bigskip

\noindent
{\large{{\bf Acknowledgements}}}

\bigskip

This work has been partially supported by DGICYT (Project  PB94--1115) 
from the
Ministerio de Educaci\'on y Ciencia de Espa\~na. 

\bigskip



\end{document}